\documentclass[aps,pra,reprint,groupedaddress,showpacs,showkeys]{revtex4-1}
\usepackage{graphicx}

 \usepackage{hyperref}
 \hypersetup{
    bookmarks=true,         % show bookmarks bar?
    unicode=false,          % non-Latin characters in Acrobat's bookmarks
    pdftoolbar=true,        % show Acrobat's toolbar?
    pdfmenubar=true,        % show Acrobat's menu?
    pdffitwindow=false,     % window fit to page when opened
    pdfstartview={FitH},    % fits the width of the page to the window
    pdftitle={YbYb*Chaos},   % title
    pdfauthor={D. G. Green},     % author
    pdfsubject={YbYb*Chaos},   % subject of the document
    pdfcreator={D. G. Green},   % creator of the document
    pdfproducer={Producer}, % producer of the document
    pdfkeywords={keywords}, % list of keywords
    pdfnewwindow=true,      % links in new window
    colorlinks=true,       % false: boxed links; true: colored links
    linkcolor=blue,          % color of internal links
    citecolor=blue,        % color of links to bibliography
    filecolor=magenta,      % color of file links
    urlcolor=blue           % color of external links
}

\begin{document}
\title{Quantum chaos in ultracold collisions between Yb($^1$S$_0$) and Yb($^3$P$_2$)}
\author{Dermot G. Green}
\altaffiliation[Present address:~]{Centre for Theoretical Atomic, Molecular and
Optical Physics, School of Mathematics and Physics, Queen's University Belfast,
Belfast BT7 1NN, Northern Ireland, United Kingdom}
\email{dermot.green@balliol.oxon.org}
\author{Christophe L. Vaillant}
\email{c.l.j.j.vaillant@durham.ac.uk}
\author{Matthew D. Frye}
\author{Masato Morita}
\author{Jeremy M. Hutson}
\email{j.m.hutson@durham.ac.uk}
\affiliation{{Joint Quantum Centre (JQC) Durham-Newcastle}, Department of Chemistry,
Durham University, South Road, Durham, DH1 3LE, United Kingdom.}
\date{\today}

%======================================
%============  ABSTRACT ===============
%======================================
\begin{abstract}
We calculate and analyze Feshbach resonance spectra for ultracold Yb($^1$S$_0$)
+ Yb($^3$P$_2$) collisions as a function of an interatomic potential scaling
factor $\lambda$ and external magnetic field. We show that, at zero field, the
resonances are distributed randomly in $\lambda$, but that signatures of
quantum chaos emerge as a field is applied. The random zero-field distribution
arises from superposition of structured spectra associated with individual
total angular momenta. In addition, we show that the resonances in magnetic
field in the experimentally accessible range 400 to 2000~G are chaotically
distributed, with strong level repulsion that is characteristic of quantum
chaos.
\end{abstract}
 %====================================================
\pacs{34.50.-s, 05.45.Mt}
\maketitle
 %====================================================
\section{Introduction}
Ultracold collisions involving the lanthanides Er and Dy in magnetic fields
exhibit dense Feshbach resonance spectra that show strong signatures of quantum
chaos \cite{Frisch:2014, Maier:ChaosErDy:2015,Maier:universal:2015}. The
density of the spectra results from large reduced masses that produce a large
number of bound levels. The complexity arises from anisotropic interactions,
which couple different end-over-end angular momenta of the colliding pair
\cite{Reid:1969, Krems:atoms:2004}, combined with magnetic fields, which couple
different total angular momenta. Chaotic behavior is likely to make the
assignment of quantum numbers to individual resonances and prediction of their
positions impossible. However, spectra of this type are amenable to statistical
analysis \cite{Wigner:statisticalnuclear:1951} that can yield physical insight
on the system \cite{Guhr:1998}, identifying the presence of good quantum
numbers or strong mixing.

Statistical analysis of complex spectra and sets of levels has been applied to
a plethora of physical systems. These include nuclear energy
levels~\cite{Brody:rmtreview:1981}, spectra of complex
atoms~\cite{Rosenzweig:levelrepulsion:1960, Flambaum:atomchaos:1994} and ions
\cite{Flambaum:manyechaos:2015} and Rydberg spectra of hydrogen atoms in large
magnetic fields~\cite{Friedrich:rydbergfield:1989}. The statistics that are
most commonly studied include the distribution of nearest-neighbor level
spacings and the level number variance \cite{Guhr:1998, Mehta:rmt:1991}. The
nearest-neighbor spacing (NNS) distribution of a randomly distributed set of
levels is of Poisson type, while that of a chaotically distributed set is of
Wigner-Dyson type. The Wigner-Dyson distribution exhibits strong level
repulsion, i.e., vanishingly small probabilities of finding levels that
coincide. The Feshbach resonance spectra for Er+Er and Dy+Dy show statistics
that indicate a considerable degree of chaos \cite{Frisch:2014,
Maier:ChaosErDy:2015, Mur-Petit:2015}, which for Dy increases steadily with
magnetic field \cite{Maier:ChaosErDy:2015}.

The appearance of chaos in ultracold collision systems has important
consequences for their properties. Chaos implies full redistribution of energy
between all degrees of freedom. It is likely to result in long-lived 2-body
collisions, which in turn can produce 3-body losses \cite{Mayle:2012}. It is
therefore very important to delineate the circumstances in which chaos arises.
Er+Er and Dy+Dy are very complex systems involving many different electronic
states. By contrast, ultracold collisions in the simpler system Li+Er
\cite{Gonzalez-Martinez:2015} have recently been shown {\em not} to exhibit
chaos. In this paper, we calculate and analyze the spectrum of Feshbach
resonance positions in ultracold collisions between bosonic ground-state
Yb($^1$S$_0$) and metastable Yb($^3$P$_2$) ytterbium atoms. We show that, even
in this remarkably simple system, application of a magnetic field induces a
transition to strongly chaotic statistics.

Yb($^1$S$_0$)+Yb($^3$P$_2$) is of interest for applications in quantum
information processing \cite{Gorshkov:quantumregisters:2009} and quantum
computing \cite{Shibata:qip:2009, Daley:qip:2008}. Takahashi and coworkers have
measured Feshbach resonances in this system \cite{Kato:ybresonances:2013,
Taie:FermYb2:2015}, and we discuss how the signatures of quantum chaos could be
observed with current experimental capabilities.

\section{Calculation of near-threshold bound states}
Yb($^1$S$_0$)+Yb($^3$P$_2$) is a particularly simple case of atom-atom
interactions with strong anisotropy. In a spin-orbit-free representation, there
are only four electronic states arising from the interaction, of which two
($^3\Sigma_g$ and $^3\Pi_g$) contribute to s-wave scattering. When spin-orbit
coupling is included, there are three Born-Oppenheimer curves that correlate
with the $^1$S$_0$+$^3$P$_2$ threshold. This contrasts with 49 and 81 curves
for the $^3$H$_6$ and $^5$I$_8$ states of the submerged f-shell atoms Er and
Dy.

Zero-energy Feshbach resonances occur when bound or quasibound states cross the
energy threshold of the entrance channel as a function of a parameter such as
magnetic field \cite{Chin:RMP:2010}. In this work we perform coupled-channel
calculations to obtain the positions of near-threshold bound levels as a
function of either magnetic field $B$ or a constant $\lambda$ that scales the
interatomic interaction potential $V\to \lambda V$. Such a potential scaling
factor, previously used to explore the sensitivity of coupled-channel
calculations to uncertainties in the potential~\cite{Cvitas:li3:2007,
Wallis:MgNH:2009, Wallis:LiNH:2011}, is used here to sample different
Hamiltonians while retaining a realistic model of the system.

\begin{figure}[t!]
\includegraphics*[width=0.42\textwidth]{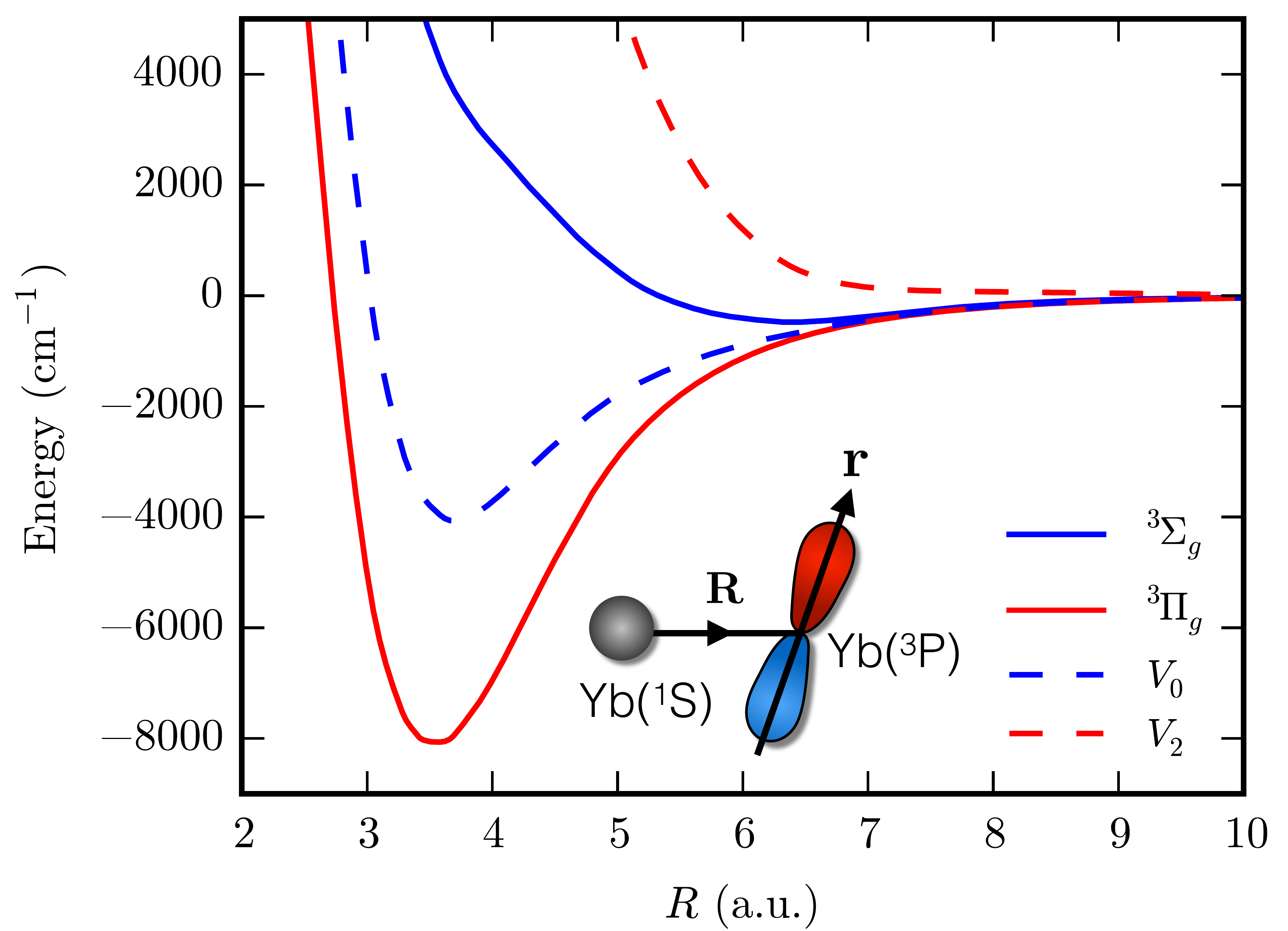}
\caption{
Interatomic potentials for Yb($^1$S)+Yb($^3$P). $\Sigma$ and $\Pi$
Born-Oppenheimer potentials calculated from \cite{Wang:Yb:1998} (red lines).
Isotropic $V_0(R)$ and anisotropic $V_2(R)$ Legendre expansion coefficients,
as described in the text (blue lines). } \label{fig:pots}
\end{figure}

We solve the Schr\"odinger equation for bound states or scattering in
coupled-channel form. We use the atom-atom Hamiltonian described in Ref.\
\cite{Gonzalez-Martinez:LiYb:2013}, except that in the present case
Yb($^3$P$_j$) interacts with a structureless partner. The interaction potential
$\hat{V}$ can be written as the Legendre expansion $\hat{V}\left({\bf
R},\hat{\bf r}\right)=\sum_{k=0,2}V_k(R) P_k\left(\hat{{\bf R}}\cdot\hat{\bf
r}\right)$ \cite{Callaway:1965, Reid:1969}, where ${\bf R}$ is the internuclear
separation vector and $\hat{\bf r}$ is a unit vector describing the position of
the Yb 6p electron. The expansion coefficients are $V_0 =
\left(V_{\Sigma}+2V_{\Pi}\right)/3$ and $V_2 =
{5}\left(V_{\Sigma}-V_{\Pi}\right)/3$ \cite{Aquilanti:1980,Krems:atoms:2004},
where $V_{\Sigma}$ and $V_{\Pi}$ are the $^3\Sigma_g$ and $^3\Pi_g$
Born-Oppenheimer potentials. Figure \ref{fig:pots} shows the $^3\Sigma_g$ and
$^3\Pi_g$ potentials of Ref.~\cite{Wang:Yb:1998}, together with $V_0$ and
$V_2$. Physically, the anisotropy is due to the 6p valence electron of
[Xe]4f$^{14}$6s6pYb($^3$P). As a result, the anisotropy in this system is much
larger than in the Er+Er and Dy+Dy systems, which involve f-shell electrons
submerged beneath a closed 6s shell \cite{Frisch:2014,
Maier:ChaosErDy:2015}. We extrapolate the potentials at long range with the
dispersion form $-C_6/R^6$ \cite{Stone:1996}, using calculated dispersion
coefficients of 2999 and 2649 $E_{\rm h}a_0^6$ \cite{Porsev:2014} for the
$^3\Sigma_g$ and $^3\Pi_g$ states respectively. The spin-orbit interaction is taken to be
independent of ${\bf R}$, with a coupling constant that gives the correct
splitting between the {$^3{\rm P}_2$ and $^3{\rm P}_1$} states
\cite{Meggers:1978}.

At zero field the total angular momentum is a good quantum number, due to the
isotropy of free space. In the absence of a field, we use the space-fixed total
angular momentum basis set $|(ls)j L J M_J\rangle$ \cite{Reid:1969}. Here the
atomic orbital and spin angular momenta $l$ and $s$ couple to give a resultant
$j$, which then couples to the end-over-end angular momentum $L$ to give the
total angular momentum $J$. At finite magnetic field we use the partially
uncoupled basis set $|(ls)jm_jLM_L\rangle$, where $m_j$ and $M_L$ are the
projections of $j$ and $L$ onto the field axis, respectively
\cite{Gonzalez-Martinez:LiYb:2013}. We include values of $L$ up to
$L_{\mathrm{max}}=22$, for which the pattern of the Feshbach resonance spectrum
is converged. Increasing the value of $L_{\mathrm{max}}$
introduces additional bound states, but they are very weakly coupled to the
entrance channel.

The coupled equations for atom-atom scattering are solved using the MOLSCAT
package \cite{molscat:v14}, modified to handle magnetic fields
\cite{Gonzalez-Martinez:2007} and P-state atoms
\cite{Gonzalez-Martinez:LiYb:2013}. Bound states are located using the FIELD
package \cite{Hutson:field:2011}, which solves the coupled-channel equations
subject to bound-state boundary conditions, using the methods of Ref.\
\cite{Hutson:CPC:1994}, to locate the magnetic fields at which bound states
exist with a specified binding energy.

In this work, we consider resonances in s-wave collisions of Yb($^3$P$_2$) in
its $m_j=-2$ state. This is the lowest component of the $j=2$ manifold.
Inelastic decays to the $j=0$ and $j=1$ manifolds in 2-body collisions with
Yb($^1$S) are slow, with a decay rate that has been measured to have an upper
bound of $10^{-13}$ cm$^{3}~$s$^{-1}$ at fields below 1~G \cite{Uetake:2012}.
We have performed coupled-channel calculations of the inelastic rate over the
range 0 to 2000~G, and find the background rate to be significantly smaller
than this bound, on the order of $10^{-17}$ cm$^{3}$ s$^{-1}$. The slow 2-body
decay makes experiments on 3-body losses in this system viable.

Using the FIELD package at the energy of the lowest threshold produces a list
of fields at which zero-energy Feshbach resonances occur
\cite{Suleimanov:2011}. For the present work we extended the FIELD package to
converge on levels (and thus resonance positions) as a function of potential
scaling factor $\lambda$ as well as magnetic field. In order to locate
resonances at the $j=2$, $m_j=-2$ threshold, basis functions for $j=0$ and 1
were omitted, corresponding to neglect of the slow inelastic decays considered
above. We expect this approximation to have no significant effect on level
statistics.

\begin{figure*}[t!]
\includegraphics*[width=1.0\textwidth]{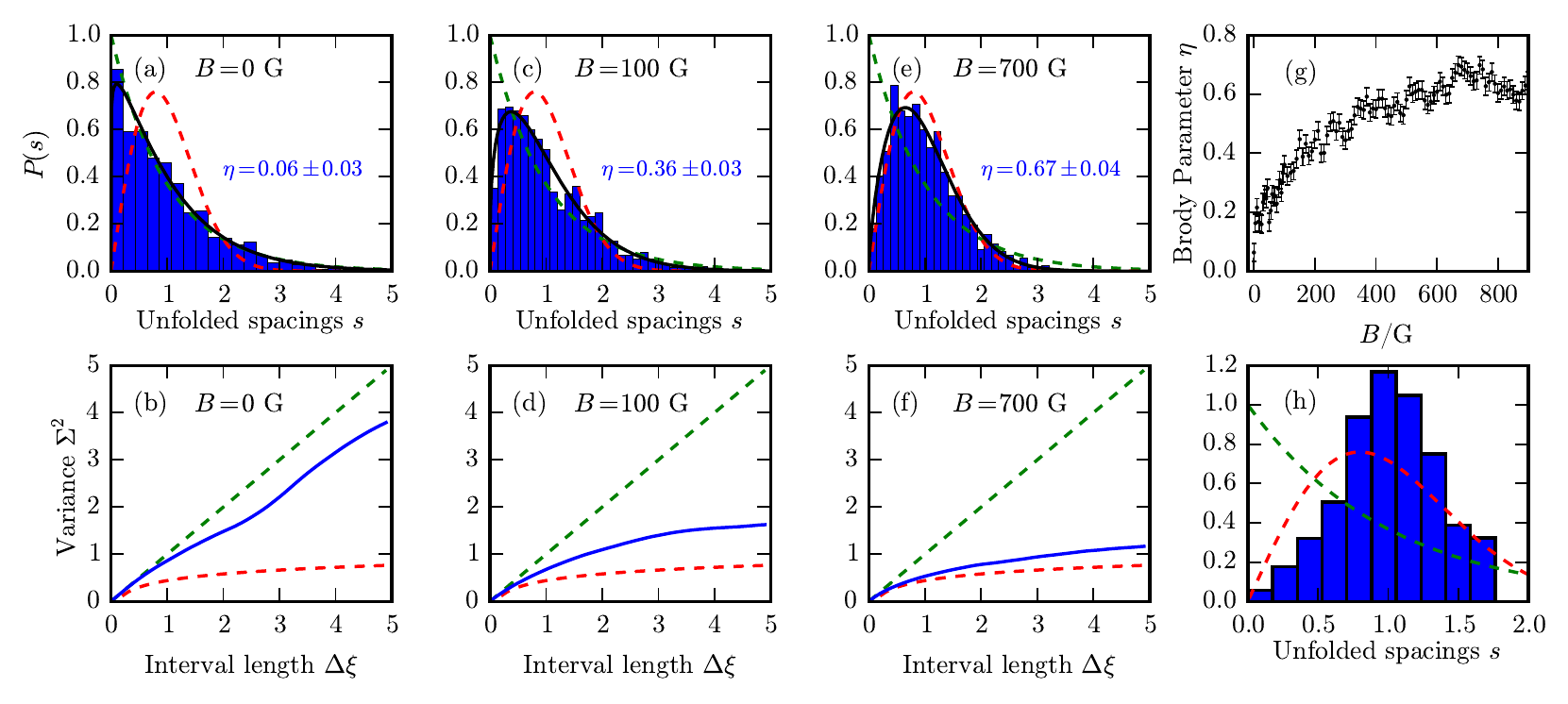}
\caption{Statistical analysis of Feshbach resonance positions with respect to
potential scaling factor $\lambda$. (a), (c) and (e) show the NNS distributions
$P(s)$: coupled-channel results (blue histograms); fitted Brody distributions
(black lines, with the corresponding Brody parameters stated); Poisson and
Wigner-Dyson distributions (green and red dashed lines, respectively).  (b),
(d) and (f) show the corresponding number variances $\Sigma^2 (\Delta \xi)$:
coupled-channel results (blue solid lines); Poisson and GOE results (green and
red dashed lines respectively). (g) Calculated Brody parameter as a function of
magnetic field. (h) NNS distribution for individual blocks of total angular
momentum $J$ in the absence of the magnetic field, averaged over
$J=2,\dots,20$. \label{fig:label}}
\end{figure*}

\section{Statistical Analysis}
We analyze sets of levels (or resonance positions) through two commonly used
statistics: the nearest-neighbor spacing (NNS) distribution, which is the
probability density $P(s)$ of two neighboring levels having the spacing $s$,
and the variance in the number of levels in a given energy range. These
statistics probe short-range correlations (on the order of a few mean level
spacings) and longer-range correlations, respectively \cite{Mehta:rmt:1991,
Guhr:1998}.

The NNS distribution and number variance must be calculated from a set of
levels on a dimensionless scale with unit local number density: the `unfolded'
scale. To obtain the unfolded levels from the calculated ones ${X_i}$, we first
construct the staircase function $S(X)=\sum_i \Theta(X - X_i)$, where $\Theta$
is the Heaviside function. $X$ is commonly the energy but here is either $B$ or
$\lambda$. We then fit a low-order polynomial $\xi(X)$ to the smoothly varying
average density, isolating the fluctuations that are of interest, and obtain
the unfolded levels $\xi_i$ by mapping $X_i\to \xi_i=\xi(X_i)$
\cite{Guhr:1998} \footnote{In general, $S(X)$ can be decomposed as
$S(X)=\xi(X)+S_{\mathrm{fl}}(X)$, where $\xi(X)$ is a smooth part given by the
cumulative mean level density, and $S_{\mathrm{fl}}(X)$ describes fluctuations
about this average. The unfolding procedure rescales the staircase function
$S(X)\to S(\xi)=\xi+S_{\rm fl}(\xi)$, i.e., to unit average density, isolating
the fluctuating part that is of interest.}.

The NNS distribution is commonly used to distinguish between regular and
chaotic systems. For an uncorrelated (random) spectrum, the NNSs on the
unfolded scale $s_i= \xi_{i+1} - \xi_{i}$ are distributed according to Poisson
statistics $P_\mathrm{P}(s)= \exp({-s})$ \cite{Guhr:1998,Mehta:rmt:1991}.
By contrast, for a chaotic system the distribution is well approximated by the
Wigner-Dyson form $P_\mathrm{WD}(s) = (\pi s/2) \exp(-\pi s^2 / 4)$
\cite{Guhr:1998, Mehta:rmt:1991}, with strong level repulsion. This
distribution is an approximation to the prediction of the Gaussian Orthogonal
Ensemble (GOE)~\cite{Guhr:1998}, which the Bohigas-Giannoni-Schmit
conjecture~\cite{Bohigas:1984} suggests is the appropriate RMT
model for a quantum system that is chaotic in the classical limit.

Physical systems rarely conform to either of these special cases. Following
recent practice in the cold-matter literature \cite{Frisch:2014,
Maier:ChaosErDy:2015, Jachymski:ChaoticScat:2015}, we interpolate between the
Poisson and Wigner-Dyson cases using the Brody ansatz $P^{(\eta)}_\mathrm{B}
(s) = c_{\eta}(1+\eta) s^\eta \exp\left(-{c_\eta s^{\eta+1}}\right)$, where
$c_\eta= \Gamma\left[(\eta+2)/(\eta+1)\right]^{\eta+1}$
\cite{Brody:1973}. We note that other methods of interpolating between
the Poisson and Wigner-Dyson nearest-neighbor distributions exist, including
rigorous semiclassical expressions \cite{Berry:semiclassical:1984}. The
`Brody parameter' $\eta$ takes values between zero (Poisson distribution) and
unity (Wigner-Dyson distribution). We calculate $\eta$ by maximum likelihood
estimation~\cite{Barlow:1989}, maximizing the log-likelihood function $l(\eta) =
\sum_i \ln P_{\mathrm B}^{(\eta)} (s_i)$ with respect to $\eta$. The
uncertainty on $\eta$ is thus the standard deviation $\sigma= \left( -
\mathrm{d}^2 l/\mathrm{d} \eta^2\right)^{-1/2}$.

The second statistic that we consider, to probe long-range correlations, is the
level number variance $\Sigma^2$. This is defined as $ {\Sigma^2 (\Delta\xi)=
\langle \hat{S}^2 (\Delta\xi, \xi) \rangle - \langle \hat{S} (\Delta\xi,
\xi)\rangle^2}, $ where $\hat{S} (\Delta\xi, \xi)$ counts the number of levels
in the interval $[\xi, \xi + \Delta\xi]$ and the average is taken over the
starting values $\xi$ \cite{Guhr:1998,Mehta:rmt:1991}. For a randomly
distributed (Poisson) set it is $\Sigma^2(\Delta\xi)= \Delta\xi$, whereas for a
Hamiltonian belonging to the GOE it is ${\Sigma^2(\Delta\xi) = 2\pi^{-2} \left[
\ln(2\pi \Delta\xi) + \gamma + 1 - \pi^{2}/{8} \right] +
\mathcal{O}(\Delta\xi^{-1})}$, where $\gamma= 0.5772 \dots$ is Euler's constant
\cite{Guhr:1998, Mehta:rmt:1991}.

\begin{figure*}[t!]
\includegraphics[width=1\textwidth]{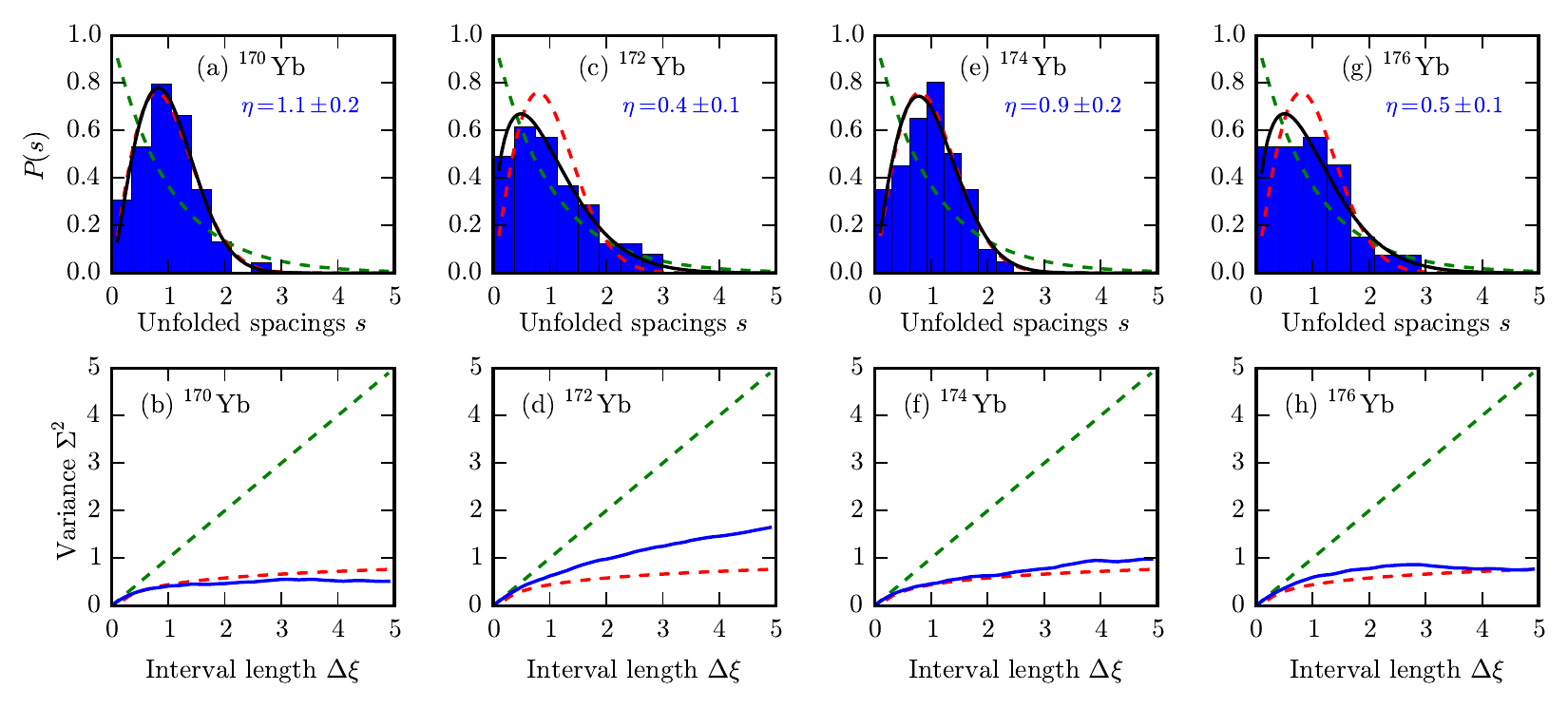}%
\caption{\label{fig:isotopeaverage} Statistical analysis of Feshbach resonance
positions with respect to magnetic field for different isotopes of Yb. Upper
panels show NNS distributions $P(s)$: coupled-channel calculations (blue
histograms); fitted Brody distributions (black lines, with the corresponding
Brody parameters stated); Poisson and Wigner-Dyson distributions (green and
red dashed lines respectively). Lower panels show the corresponding number
variances $\Sigma^2 (\Delta \xi)$: coupled-channel calculations (blue solid
lines); Poisson and GOE results (green and red dashed lines, respectively). }
\end{figure*}

\section{Results and discussion}
Figure \ref{fig:label}(a)--(f) shows the NNS distribution and number variance
for a sequence of 1000 resonance positions, calculated with respect to
$\lambda$ on the range [0.9,1.13], in external magnetic fields of 0~G, 100~G
and 700~G. In the absence of a field, the NNS distribution and the number
variance are close to those expected for Poisson statistics, with a Brody
parameter $\eta=0.06\pm 0.03$. However, application of a magnetic field induces
a clear transition towards chaotic statistics. Fig.~\ref{fig:label}(g) shows
the Brody parameter $\eta$ as a function of field $B$: it rises steadily from
close to zero at $B=0$ to a value around 0.6 at fields above 500~G. The
high-field value is comparable to that observed experimentally for Er and Dy
\cite{Maier:ChaosErDy:2015}. The number variance also changes steadily from
near-Poissonian to chaotic behavior as the field increases, following the GOE
prediction at high field more strongly than for Er and Dy.

Let us consider further the result at zero field, where the total angular
momentum $J$ is a good quantum number. In Fig.~\ref{fig:label}(h) we show the
NNS distribution for individual Hamiltonian blocks of a given total angular
momentum $J$, averaged over values of $J=2,\dots,20$ to obtain improved
statistics \footnote{The average is performed by first calculating the
unfolded spacings for each $J$, before combining the sets and normalizing the
resulting histogram.}. Although this superficially resembles a Wigner-Dyson
distribution, except that there is a cutoff at large spacing, the levels
associated with individual blocks of the total Hamiltonian are in fact highly
structured. They correspond to the superposition of nearly independent
sequences for $|\Omega|=0$, 1 and 2, where $\Omega$ is the projection of $J$
onto the interatomic axis \footnote{The cutoff at approximately $s=1.8$ in
Fig.~\ref{fig:label}(h) is consistent with the vibrational spacing for the
$|\Omega|=2$ potential at dissociation, calculated with respect to $\lambda$.
This is the deepest of the three potentials, and is equivalent to the $^3\Pi_g$
potential.}. It is evident that the Poisson statistics exhibited by the full
spectrum at zero field result from superposition of these structured spectra.

Thus far we have considered the distribution of resonances with respect to an
interatomic potential scaling factor. We now consider the distribution of
Feshbach resonances with respect to magnetic field, for homonuclear collisions
involving the four most abundant bosonic isotopes of Yb. The typical density of
resonances is $\sim 0.05$ G$^{-1}$. This is comparable to that found in Cs
\cite{Berninger:Cs2:2013} and Li+Er \cite{Gonzalez-Martinez:2015}, but less
than that observed in the Er and Dy systems, where it can be as large as $\sim$
4 G$^{-1}$ (for bosonic isotopes) \cite{Frisch:2014, Maier:ChaosErDy:2015}.

Figure \ref{fig:isotopeaverage} shows the NNS distributions and number
variances for $^{170}$Yb, $^{172}$Yb, $^{174}$Yb and $^{176}$Yb in the field
range 400 to 2000~G. The statistics show strong signatures of chaos in each
case, with Brody parameters ranging from 0.5 to about 1 and number variances
much closer to the GOE predictions than to Poisson statistics. We emphasize
that the statistics depend on the potential scaling factor as well as the
isotopic mass, so the results in Fig.\ \ref{fig:isotopeaverage} are
representative of typical behavior, rather than specific predictions for
individual isotopes. Signatures of chaos emerge at somewhat different fields
for different cases, but are always strongly present for fields over 600~G.
These signatures will be observable if current experiments on Feshbach
resonances in Yb($^1$S$_0$)+Yb($^3$P$_2$) \cite{Kato:ybresonances:2013,
Taie:FermYb2:2015} can be extended to suitable magnetic fields.

The results in Figs. 2 and 3 show that a large number of electronic states is
not required for signatures of chaos to emerge in ultracold collisions, as may
have been expected from the Er and Dy examples. We conclude that chaos in
Yb+Yb* emerges as a result of the combination of strongly anisotropic
interactions and magnetic field, consistent with the findings for
Dy+Dy~\cite{Maier:ChaosErDy:2015}. As a counterexample, we have analyzed the
Feshbach resonance positions in Cs($^2$S)+Cs($^2$S) collisions in magnetic
field~\cite{Berninger:Cs2:2013}, where there are two electronic states but only
very weak anisotropy. We find no deviations from Poisson statistics for Cs.

% ================================================
% ====== Summary/outlook ============================
% ================================================
\section{Conclusions}
We have calculated and statistically analyzed the
positions of Feshbach resonances for collisions of ground-state and metastable
Yb. This is one of the simplest possible cases of atom-atom interactions with
strong anisotropy. Even in this remarkably simple system, the application of an
external magnetic field induces a transition from random (Poisson) statistics
at zero field to chaotic statistics at high field. This suggests that chaos is
likely to be widespread in ultracold collisions, which will have important
consequences for the lifetimes of ultracold species. We predict that the
positions of magnetically tunable Feshbach resonances for the four most
abundant bosonic Yb isotopes will exhibit strong signatures of quantum chaos at
high magnetic fields. These signatures could be observed in experiments within
reach of current technology.

\begin{acknowledgments}
The authors thank C.~Ruth Le Sueur, Maykel L.~Gonz\'{a}lez-Mart\'{\i}nez and
Paul S.~Julienne for valuable discussions. This work was supported by the
Engineering and Physical Sciences Research Council under grant number
EP/I012044/1. The data presented in this paper are available online~\cite{Green:data:2015}.

\end{acknowledgments}

\bibliography{../../all,Yb2_chaos}

\end{document}